\documentclass[aastex, twocolumn]{emulateapj}
\usepackage{amsmath}
\usepackage{longtable}
\usepackage{array}
\usepackage{hyperref}
\hypersetup{colorlinks=true, linkcolor=blue, citecolor=blue, urlcolor=blue}
\usepackage{epstopdf}

\shorttitle{Chromospherically Active Stars in the RAVE Survey}
\shortauthors{\v Zerjal et al.}

\begin{document}
\title{Chromospherically Active Stars in the RAVE Survey. I. The Catalogue}

\author{
M. \v Zerjal\altaffilmark{1}, 
T. Zwitter\altaffilmark{1,2}, 
G. Matijevi\v c\altaffilmark{3}, 
K. G. Strassmeier\altaffilmark{4},
O. Bienaym\'{e}\altaffilmark{5},
J. Bland-Hawthorn\altaffilmark{6},
C. Boeche\altaffilmark{7},
K.~C. Freeman\altaffilmark{8},
E.~K. Grebel\altaffilmark{7},
G. Kordopatis\altaffilmark{9},
U. Munari\altaffilmark{10},
J. F. Navarro\altaffilmark{11},
Q. A. Parker\altaffilmark{12, 13, 14},
W. Reid\altaffilmark{12, 13},
G. Seabroke\altaffilmark{15},
A. Siviero\altaffilmark{4, 16},
M. Steinmetz\altaffilmark{4},
R.~F.~G. Wyse\altaffilmark{17}
}

\altaffiltext{1}{Faculty of Mathematics and Physics, University of Ljubljana, Jadranska 19, 1000 Ljubljana, Slovenia; \email{marusa.zerjal@fmf.uni-lj.si}}
\altaffiltext{2}{Center of Excellence SPACE-SI, A\v sker\v ceva cesta 12, 1000 Ljubljana, Slovenia}
\altaffiltext{3}{Dept. of Astronomy and Astrophysics, Villanova University, 800 E Lancaster Ave, Villanova, PA 19085, USA}
\altaffiltext{4}{Leibniz-Institut f\"ur Astrophysik Potsdam (AIP), An der Sternwarte 16, D-14482, Potsdam, Germany}
\altaffiltext{5}{Observatoire astronomique de Strasbourg, Universit\'{e} de Strasbourg, CNRS, 11 rue de l'Universit\'{e}, 67000 Strasbourg, France}
\altaffiltext{6}{Sydney Institute for Astronomy, School of Physics A28, Sydney, Australia, NSW 2006}
\altaffiltext{7}{Astronomisches Rechen-Institut, Zentrum f\"ur Astronomie der Universit\"at Heidelberg, M\"onchhofstr. 12-14, D-69120 Heidelberg, Germany}
\altaffiltext{8}{Research School of Astronomy and Astrophysics, Australia National University, Weston Creek, Canberra, ACT 2611, Australia}
\altaffiltext{9}{Institute of Astronomy, Cambridge University, Madingley Road, Cambridge CB3 0HA, UK}
\altaffiltext{10}{INAF Osservatorio Astronomico di Padova, 36012 Asiago, Italy}
\altaffiltext{11}{Senior CIfAR Fellow. Department of Physics and Astronomy, University of Victoria, Victoria BC, Canada V8P 5C2}
\altaffiltext{12}{Department of Physics \& Astronomy, Macquarie University, Sydney, NSW 2109 Australia}
\altaffiltext{13}{Research Centre for Astronomy, Astrophysics and Astrophotonics, Macquarie University, Sydney, NSW 2109 Australia}
\altaffiltext{14}{Australian Astronomical Observatory, PO Box 296 Epping, NSW 1710, Australia}
\altaffiltext{15}{Mullard Space Science Laboratory, University College London, Holmbury St Mary, Dorking, RH5 6NT, UK}
\altaffiltext{16}{Department of Physics and Astronomy, Padova University, Vicolo dell’Osservatorio 2, I-35122 Padova, Italy}
\altaffiltext{17}{Johns Hopkins University, Homewood Campus, 3400 N Charles Street, Baltimore, MD 21218, USA}

\begin{abstract}
RAVE, the unbiased magnitude limited survey of the southern sky stars, contained 456,676 medium-resolution spectra at the time of our analysis. Spectra cover the Ca~II~IRT range which is a known indicator of chromospheric activity. Our previous work \citep{2012ApJS..200...14M} classified all spectra using locally linear embedding. It identified 53,347 cases with a suggested emission component in calcium lines.
Here we use a spectral subtraction technique to measure the properties of this emission. Synthetic templates are replaced by the observed spectra of non-active stars to bypass the difficult computations of non-LTE profiles of the line cores and stellar parameter dependence.
We derive both the equivalent width of the excess emission for each calcium line on a $5\;\text{\AA}$ wide interval and their sum $\mathrm{EW_{IRT}}$ for $\sim$~44,000 candidate active dwarf stars with S/N~$>20$ and with no respect to the source of their emission flux. From these $\sim$~14,000 show a detectable chromospheric flux with at least $2\, \sigma$ confidence level. Our set of active stars vastly enlarges previously known samples.
Atmospheric parameters and in some cases radial velocities of active stars derived from automatic pipeline suffer from systematic shifts due to their shallower calcium lines. We re-estimate the effective temperature, metallicity and radial velocities for candidate active stars.
The overall distribution of activity levels shows a bimodal shape, with the first peak coinciding with non-active stars and the second with the pre main-sequence cases.
The catalogue will be publicly available with the next RAVE public data releases.
\end{abstract}

\section{Introduction}
RAVE (RAdial Velocity Experiment, \citealp{2006AJ....132.1645S, 2008AJ....136..421Z, 2011AJ....141..187S, kordopatis_dr4}) is an ongoing spectroscopic southern sky survey with unbiased magnitude limited selection of stars ($9 < I < 12$). The aim of the survey is a determination of radial velocities and an estimation of atmospheric parameters (effective temperature, surface gravity and metallicity) of Galactic stellar populations. 
The typical signal-to-noise ratio of the measured spectra is $\sim 40$, their typical resolving power is in mid-range ($R\sim7500$). All spectra in the sample are continuum normalized and shifted to zero heliocentric radial velocity. The survey covers the near-infrared range from $\sim 8410\;\text{\AA}$ to $\sim 8795\;\text{\AA}$ where the singly ionized calcium triplet ($\lambda \lambda = 8498, \; 8542, \; 8662\;\text{\AA}$) dominates the spectral shape on a broad range of atmospheric parameters with the temperature being the most important. There is practically no contribution of telluric lines in that part of the spectrum; the only significant spectral signature of the interstellar medium is a diffuse interstellar band at $8620\;\text{\AA}$ (\citealp{2008A&A...488..969M}). All spectra used in this study were obtained between 2004 April 8 and 2012 October 20.  These stars ($\sim$~350,000) have been classified with the locally linear embedding (LLE) method by \citet{2012ApJS..200...14M} in order to explore the morphology of the spectra, to reveal problematic examples and to discover peculiar stars. The sample, based on the results of the LLE, consists of $\sim 90\% - 95\%$ normal single stars. However, two large distinct groups of peculiars were found. One of them are active stars that show chromospheric emission in calcium lines ($\sim$~50,000 stars).

The phenomenon of chromospheric emission occurs in young main sequence solar-type or cooler dwarfs with convective envelopes and affects strong spectral lines with the largest impact on $\mathrm{H}\alpha$, Ca~II H~\&~K ($3969, \;3934\;\text{\AA}$) and the Ca~II IR triplet. The latter is covered by the RAVE survey. There is a strong emission in Mg~II h~\&~k cores, but it occurs at $2800\;\text{\AA}$ which is not observable from the ground. The chromospheric emission component is seen as an additional flux at the central wavelengths of strong lines. It can also be Doppler shifted with respect to the radial velocity of the stellar photosphere. Younger stars appear to be more active and since the chromospheric activity level correlates well with a diminishing stellar rotation rate through the coupling by the magnetic dynamo effect, it can be used as an age diagnostics tool for the nearest relatively young stars (0.6-4.5 Gyr, \citealp{2008ApJ...687.1264M}). The nature of chromospheric activity evolution has been questioned by \citet{2013A&A...551L...8P, 2011AJ....141..107Z, 2005A&A...431..329L}.
\citet{2008LRSP....5....2H} presents a review of recent advances in stellar chromospheric activity research and \citet{2010ARA&A..48..581S} reviews its connection to stellar ages while \citet{2009A&ARv..17..251S} discussed its connection to starspots.

The majority of emission datasets in the literature cover the Ca~II H~\&~K lines. The 'HK Project' at Mount Wilson Observatory operated from 1966 through 2003 and gathered a large collection of multiple observations of $\sim 1300$ stars over 40 years which offers an invaluable opportunity to study cyclic variations of activity \citep{1978ApJ...226..379W, 1991ApJS...76..383D, 1995ApJ...438..269B}.

Other large datasets were described by \citet{1996AJ....111..439H} ($\sim 800$ stars), \citet{2000A&AS..142..275S} ($\sim 1000$ stars), \citet{2004ApJS..152..261W} ($\sim 1200$ stars), \citet{2006AJ....132..161G} ($\sim 1700$ stars), \citet{2010ApJ...725..875I} ($\sim 2600$ main-sequence and giant stars) and \citet{2013AJ....145..140Z} (over $13,000$ F, G and K disk stars).

Chromospheric emission in Ca~II IRT, discussed below, has not been covered to a large extent yet. The line at $8542\;\text{\AA}$ was investigated for 49 stars by \citet{1979ApJS...41..481L} with a conclusion that this line is a good diagnostic of chromospheric activity. \citet{1993ApJS...86..293D} analysed Ca~II IRT lines of 45 active F2-M5 stars. \citet{1993ApJS...85..315S} used the line at $8542\;\text{\AA}$ to disentangle the activity of Pleiades cluster members. \citet{1997A&AS..124..359M} studied 146 stars brighter than $\mathrm{V}=+7.0$. \citet{2010MNRAS.407..465J} analysed 237 stars in $\mathrm{NGC\,2516}$.

Here we describe a sample of newly discovered $\sim$~44,000 candidate active stars drawn from an unbiased magnitude limited spectroscopic survey. Candidate active stars cover a wide range of activity levels, from very strong emission objects to numerous examples with only a marginal detection or no detection at all. We note that the presence of an emission component does not imply its chromospheric origin. In fact some of the most active candidate stars may well be examples of other kinds of activity, for example \citet{2005A&A...440.1105S} reports the emission component in classical T Tauri stars where the emission partly comes from accretion shocks. \citet{2010AJ....140.1758T} present an atlas of different peculiar stellar types in the RAVE wavelength range, many of them with an emission in the Ca II IRT. 
But since the chromospheric origin holds for the vast majority of cases we use this reference throughout the paper.

A detailed description of our sample of candidate active stars is given in Sec. \ref{sec.sample}. The summary of the activity analysis procedure and new parameter estimation is provided in Sec. \ref{sec.method}. In Sec. \ref{sec.activity} we comment on the distribution of stars with different activity levels, while a general discussion and conclusions are given in Sec. \ref{sec.discussion}. A description of the catalogue of chromospherically active candidate stars, which is the main result of this paper, is given in the Appendix.

\begin{figure*}
\includegraphics[width=\textwidth]{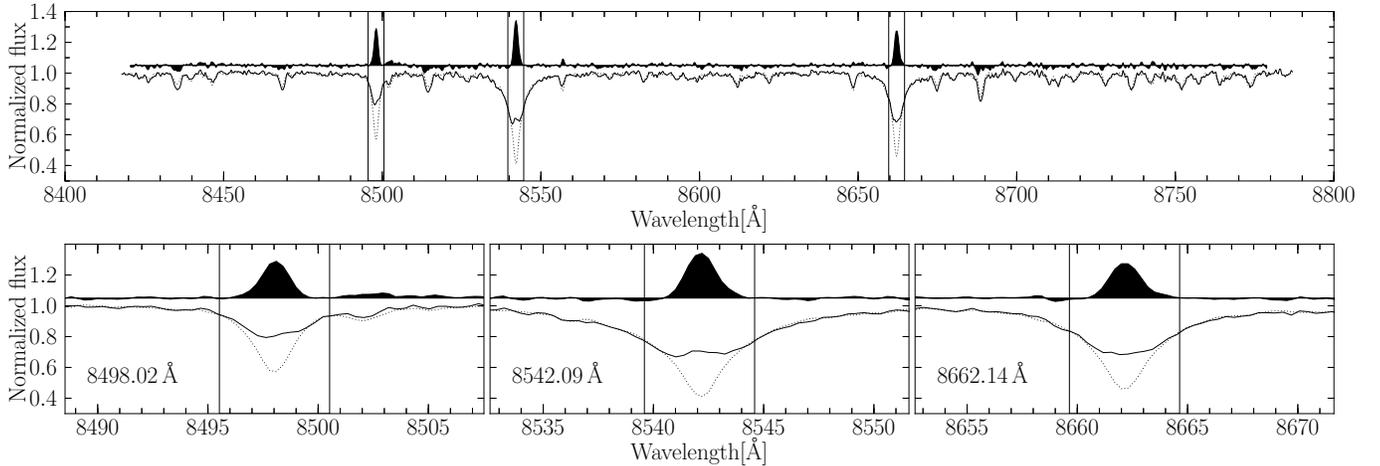}
\caption{\textit{Top panel:} An example for the RAVE spectrum of a candidate active star (HD 220054). The solid line is the spectrum itself and the dotted line is its nearest normal neighbour that is used to determine the subtracted spectrum (black area on the top). The three bands are the $5\;\text{\AA}$ wavelength ranges that are not taken into account when searching for the nearest neighbour but are used in determining stellar activity rate. \textit{Bottom panels}: Enlarged sections with calcium lines. Central wavelengths for each line are given at the left bottom corner.}
\label{fig.example_spectrum}
\end{figure*}

\begin{figure}
\includegraphics[width=\columnwidth]{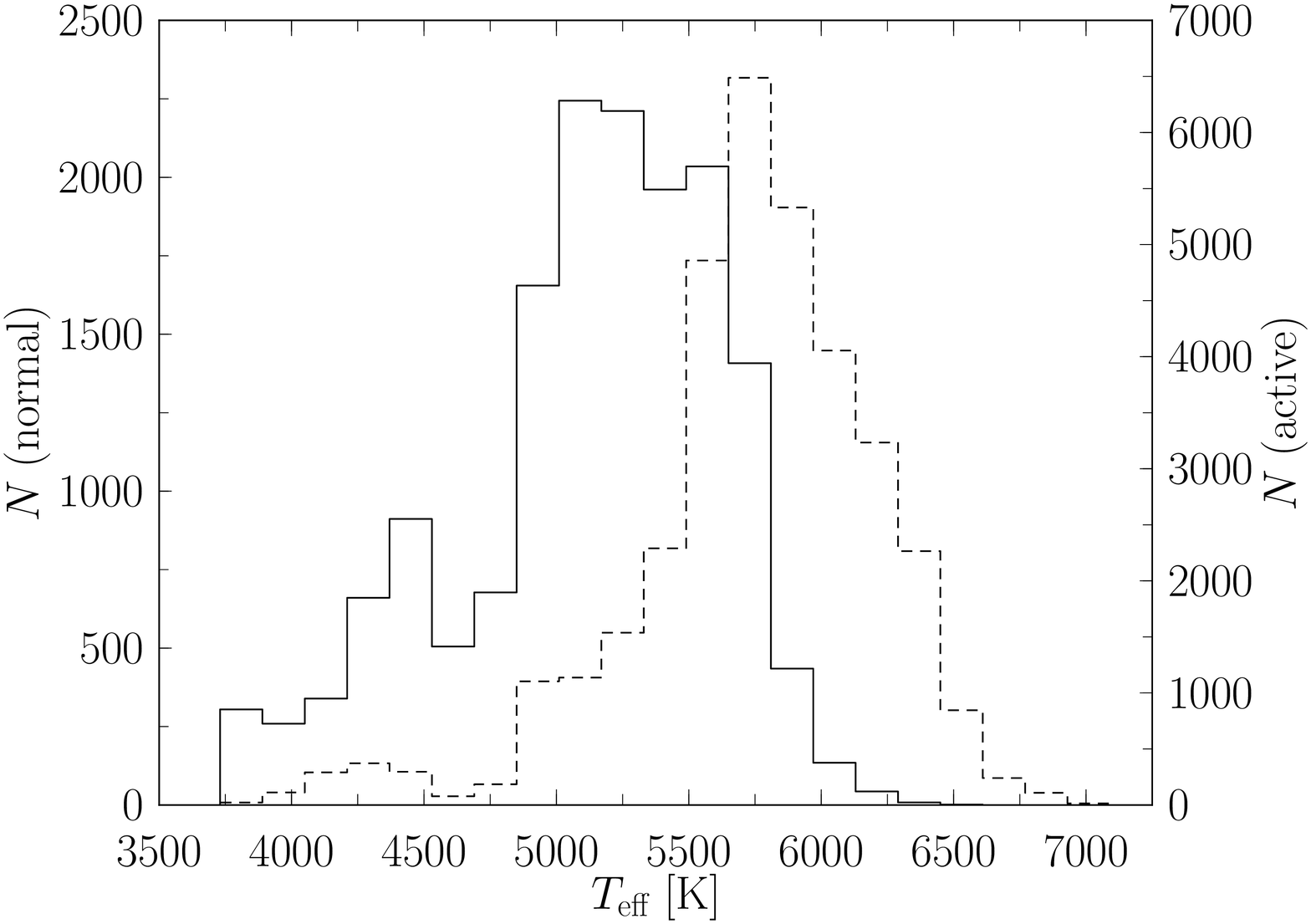}
\includegraphics[width=\columnwidth]{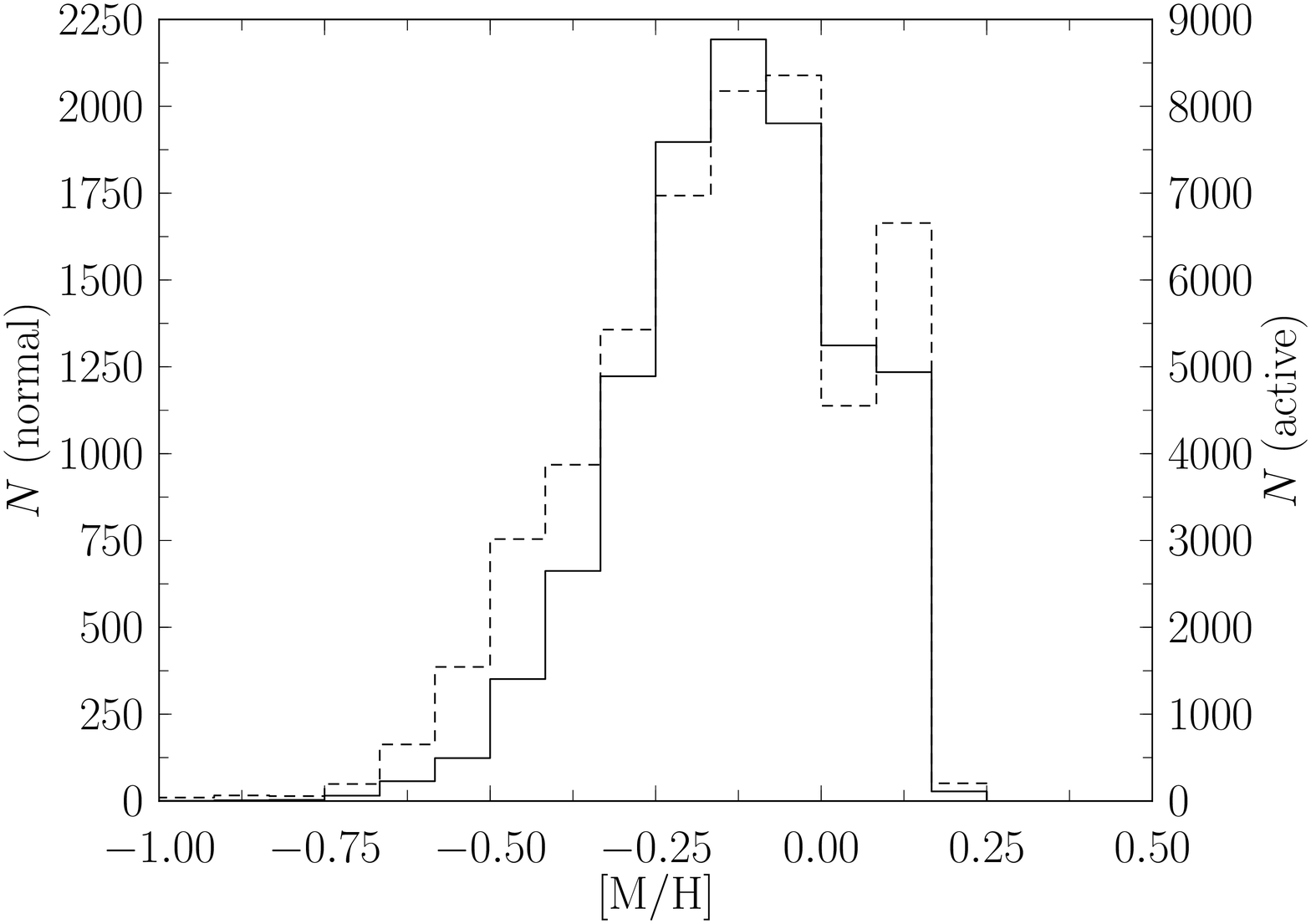}
\caption{Parameter distribution for surface temperature and metallicity. Solid line represents candidate active stars while the dashed one are inactive dwarfs (normal database). Bin heights are given on the right and left, respectively. Values for both sets on plots were re-estimated in this paper as an average of values for five closest inactive spectral tracings, in the $\chi^2$ sense.}
\label{fig.database_parameters}
\end{figure}

\section{Catalogue of candidate active stars}\label{sec.sample}

The selection of candidate active stars from the whole RAVE database of over 500,000 spectra 
is based on the classification by the LLE method. LLE is a general dimensionality reduction algorithm with a property of conserving locally linear patches of the high-dimensional manifold. So the neighbours in low   dimensional projected space are neighbours also in the original high-dimensional space. The method is executed in three separate steps. First, it finds the nearest neighbours for each data point. Then, it calculates the weights that best represent a given data point as a linear combination of its neighbours in the original space. Finally, it finds a low-dimensional space in which each data point is still best represented by the same set of weights that were calculated in the previous step. For a more
detailed description of the method see \citet{2000Science...190...2323R, 2009AJ....138.1365V}, or \citet{2012ApJS..200...14M}, and references therein.

\citet{2012ApJS..200...14M} extended the method for spectral classification purposes. Because the method is unable to handle large data samples in a reasonable amount of time, the procedure was broken down into following steps. A base set of 5000 spectra was randomly selected from the whole RAVE database and each spectrum from the set was rebinned to the common wavelength range with 1200 bins (providing 5000 data points in the 1200-dimensional wavelength space). Using the LLE method with the number of nearest neighbours set to 20, the optimal 2-dimensional space was found. While the dimensionality of the projected space might not reveal all the potentially interesting features of the underlying spectra, it is sufficient to make a distinction between various spectral groups. After the low-dimensional space was calculated, the rest of the spectra from the database of several hundred thousand objects were rebinned and projected onto this space. Due to the lack of rare peculiar morphologies in the base set (spectra in the base set were selected randomly and it is therefore unlikely that less common types were picked) the 2-dimensional projection of such objects is misleading and they are usually not grouped together properly. To overcome this problem we reduced the projected sample to a new set of ~5000 points by removing many points from the densely populated areas full of very common normal stars while preserving the outliers representing rare morphologies. This ensures the new set includes more peculiars relative to the normal stars. The newly generated sample served as the new base set for the next iteration of calculating the optimal low-dimensional space, projecting the spectra and reducing the projection. This was repeated several times to ensure that the final base set represented all spectral morphologies found in the database as uniformly as possible. Spectra in the final base set were manually classified into 11 classes. The rest of the spectra were then automatically classified by finding the 20 nearest neighbours from the base set and calculating the linear weights. This provided a quantitative measure of how much signal of each of the base neighbouring spectra is represented in each spectrum. Closest 20 neighbours were arranged in a sequence corresponding to the order of the weights, with the base spectrum having the largest weight coming first and so on.

Our catalogue of candidate active stars described in this paper includes all objects from the RAVE database that have at least one chromospherically active neighbour within its first 6 neighbours.
The number of neighbours to sufficiently describe the morphology of the spectrum is chosen as the limit where the contribution of the next successor to the information about the spectrum is negligible.
Furthermore, we kept only those objects with $S/N$ greater than 20 because under this limit the re-estimation of atmospheric parameters, radial velocity (not based on calcium lines but on weaker Fe~I and Ti~I lines) and activity levels are unreliable. We loose $\sim 18\%$ of stars with this S/N cut, but since we have a large database the overall quality of the rest of the database is better. Our final database of candidate active spectra contains $\sim$~44,000 stars that were recognized as candidates for chromospheric activity by LLE classification (\citealp{2012ApJS..200...14M}). Most of them are main sequence stars with a K spectral type. The emission flux of the sample stars continuously ranges from an undetectable amount in almost inactive stars to examples with an emission component completely filling the calcium line and even extending above the continuum level in some cases (see Fig. \ref{fig.example_spectrum}).
Our catalogue contains candidates for chromospherically active stars. However, since only one active spectrum is required within the first 6 neighbours for stars in our catalogue, other neighbours can be of any type. Consequently, the results should be treated with caution because we analysed all objects with affected calcium lines with no respect to the origin of the additional emission flux. For this reason term 'active star' refers to candidate active star within this paper further on.

In addition to the database of active stars we use another database of measured 'normal' reference spectra which are used to compute the subtracted emission flux in this paper. Single lined main sequence dwarfs without any peculiarities (our request was that at least the first 7 LLE flags indicate that there are no peculiarities), with high S/N ($>30$), temperature below $7000\;\mathrm{K}$ and $\log{g}>3.5$ are denoted as \textit{'normal stars'} within this paper for simplicity. Our database of selected normal stars is called \textit{'normal database'}; its selection procedure is given in the next chapter. 
\citet[hereafter DR4]{kordopatis_dr4}, recently computed an updated list of stellar parameters for normal stars. Their computation is based on the DEGAS and MATISSE algorithms using \citet{2011A&A...535A.106K} grid. So we decided to use the values of stellar parameters from Kordopatis for the normal database. As we explain below also the parameter values pertaining to active stars are therefore compatible with this recent re-estimation of stellar parameters. 

All RAVE pipelines are designed for the most abundant group of the spectra ($\sim 90\% - 95\%$ of the whole dataset) - i.e., stars that show no peculiarities. Radial velocities and stellar atmospheric parameters for active spectra might suffer from systematic offsets because activity makes the three calcium lines -- a dominant feature in RAVE spectra -- shallower. One of our goals is therefore a re-estimation of stellar parameters for these stars.

\begin{table}
\begin{center}
\begin{tabular}{l | c c c c}
\hline
\# & $\mathrm{N_n}$ & $T_{\mathrm{eff}}\;[\mathrm{K}]$ & S/N & N \\
\hline
1	&	7	&	$<4500$					&	30	& 406 \\
2	&	7	&	$4500 \leq T_{\mathrm{eff}}<5000$	&	40	& 832 \\
3	&	7	&	$4500 \leq T_{\mathrm{eff}}<5500$	&	40	& 3648 \\
4	&	7	&	$5500 \leq T_{\mathrm{eff}}<6000$	&	50	& 5379 \\
5	&	10	&	$6000 \leq T_{\mathrm{eff}}<7000$	&	80	& 2187 \\
\end{tabular}
\end{center}
\caption{Selection criteria for normal inactive observed spectra (see text). The database is built from five subsets and contains 12,452 objects. The second column ($\mathrm{N_n}$) gives the number of the first nearest neighbours of the spectrum that were required to be normal (LLE based selection), and S/N marks the minimal required value. N is the number of spectra in the particular subset. Most relaxed requirements are used for the coolest stars.}
\label{tab.database_selection_criteria}
\end{table}

\section{Estimation of chromospheric activity level}\label{sec.method}
The most commonly used proxies for Ca~II H~\&~K chromopheric activity are the dimensionless S and $\mathrm{R'_{HK}}$ indices. The S index was introduced by \citet{1978PASP...90..267V} as the strength of chromospheric emission in both line cores. It is proportional to the sum of fluxes in the H~\&~K lines and normalized with respect to the nearby continuum: $S \propto (\mathcal{F}_\mathrm{H}+\mathcal{F}_\mathrm{K}) / (\mathcal{F}_\mathrm{R}+\mathcal{F}_\mathrm{V})$. The full-width at half-maximum of both K and H bands with a triangular profile is $\sim 1\;\text{\AA}$; fluxes $\mathcal{F}_\mathrm{R}$ and $\mathcal{F}_\mathrm{V}$ are red and violet continuum bands from $3991-4011\;\text{\AA}$ and $3891-3911\;\text{\AA}$. The S index has a color term and includes a contribution from the stellar photosphere. The second and most commonly used activity measure is $\mathrm{R'_{HK}}$
which was introduced by \citet{1984ApJ...279..763N}. The $\mathrm{R'_{HK}}$ index is the ratio of the emission flux in the cores of calcium H~\&~K lines (the bandwidth is $1\;\text{\AA}$ for each line core) to the total bolometric flux of the star.

A possible proxy for the Ca~II IRT part is a central line depression based on the flux difference between the observed central flux level and non-LTE synthetic spectrum.
It was described by \citet{2005A&A...430..669A}. \citet{1993ApJS...85..315S} and \citet{2010MNRAS.407..465J} adopted the chromospheric flux, while \citet{1997A&AS..124..359M} used the sum of equivalent widths of the three emission components. 
The empirical flux scales for spectral types A to early M for luminosity classes I to V, including the IRT range, were first derived by \citet{1996PASP..108..313H}. An analogous measure to the $\mathrm{R'_{HK}}$ for the IRT part is given in \citet{2009MNRAS.399..888M} as $R'_{\mathrm{IRT}}=F'_{IRT}/\sigma T_{\mathrm{eff}}^4$ where $F'_{IRT}$ is the surface flux emitted by IRT calcium lines. The surface flux is derived by a conversion of the equivalent width of the emission excess. The latter is revealed by a spectral subtraction technique. The idea is to find a synthetic counterpart of the observed active spectrum and to remove the photospheric contribution by subtracting the two. The subtracted spectrum represents the chromospheric emission flux.

Disadvantages of this method are the requirement of precise values of the atmospheric parameters for the active star and the non-LTE nature of the core of the line which exactly overlaps with the expected emission flux. The synthetic spectrum should also be artificially broadened due to stellar rotation (a poorly determined quantity for RAVE spectra) and appropriately adapted to the resolving power of the instrument (to be precise, instrumental PSF, spectral purity and scattered light within the spectrograph might also play a role here).

Since the majority of stars in the RAVE database are normal stars (38\% of them are dwarfs with $\log{g} \geq 3.5$), we seized the opportunity and used observed normal high S/N spectra instead of synthetic templates to compute the subtracted spectra. Our normal database contains 12,452 observed spectra. It is large enough to cover the whole parameter space of candidate active stars - effective temperature and metallicity variations, together with gravity, possible variations of resolving power along the spectrum and rotational velocity.

The best matching templates for each of the active spectra are found by the $\chi^2$ nearest neighbour search where the cores of the calcium lines are not taken into account (see Fig.~\ref{fig.example_spectrum}).
When nearest normal neighbour is found we assume that it represents the active spectrum itself in a quiet state and with the same stellar and instrumental parameters. One may ask if the selected normal stars are completely inactive or not. We checked that their activity levels are below our detection limits, so they can serve our purpose well. The equivalent widths of the chromospheric flux in the calcium lines are computed from the subtracted spectra. 
The details of the procedure are given below.

\subsection{Normal database selection}
A large database of normal spectra assumed to represent active spectra in a completely quiet state was built
to achieve a nearly uniform coverage of the whole stellar parameter space. Since chromospheric emission occurs in main-sequence dwarfs, we set the upper temperature limit of the database to $7000\;\mathrm{K}$ and the gravity limit to $\log{g}\ge 3.5$. The lower limit in the temperature is $3500\;\mathrm{K}$ because this is the limit of synthetic grid (\citealp{2005A&A...442.1127M}) used in the RAVE DR3 pipeline \citep{2011AJ....141..187S} and there is only a small number of stars with the temperature below this limit because this survey is magnitude limited and therefore its low luminosity stars lie in a small volume in the vicinity of the Sun. Most stars with true values lower than 3500~K actually fall in the $3500$ -- $3600\;\mathrm{K}$ range for the DR3 pipeline. The determination of parameters in the DR4 pipeline (\citealp{kordopatis_dr4}) for normal spectra is based on a synthetic grid with a step of $250\;\mathrm{K}$ for spectra between 4000 and $8000\;\mathrm{K}$ and $200\;\mathrm{K}$ for spectra between 3000 and $4000\;\mathrm{K}$. The first group ranges from $0.0-5.0\;\mathrm{dex}$ with a step size of $0.5\;\mathrm{dex}$ in $\log{g}$ and from $0.0-5.5\;\mathrm{dex}$ in the second group. The metallicity ranges from $-5.0$ to $+1.0\;\mathrm{dex}$ with a step size of $0.25\;\mathrm{dex}$.
Typical internal uncertainties of the RAVE pipeline output are $170\;\mathrm{K}$ in temperature, $0.27\;\mathrm{dex}$ in $\log{g}$ and $0.16\;\mathrm{dex}$ in metallicity.

A spectrum was recognized as normal in the same manner as in the case of the selection of active stars. Due to a much larger set of normal spectra the requirements for the nearest neighbours quality and confidence were more rigorous. Details are given in Table~\ref{tab.database_selection_criteria}. We note that the density of normal stars is very low outside the main sequence where possible active outliers (pre-main sequence stars) might be found. So we relaxed the selection criteria in regions scarcely populated with normal stars. In particular we were forced to accept also observed spectra with a lower S/N ratio. In total we used 5 different subsets that were merged into the final normal database (see Table~\ref{tab.database_selection_criteria}).

We sieved the database in the next step to reduce the number of objects in extremely dense parts of the parameter subspace (surface temperature, gravity and metallicity). We divided the parameter subspace into smaller volume units and kept deleting randomly selected stars until the number density in each volume unit dropped below the predefined density. Finally a set of 12,452 normal spectra remained and was declared to be our normal database. The distribution of the effective temperature and metallicity of the stars in this database is shown in Figure \ref{fig.database_parameters} together with the parameters for active stars.

\subsection{Template spectra}
For each active spectrum we need to select a few best matching template spectra from the normal database. The match was evaluated using $\chi^2$ statistics, ignoring the cores of the calcium lines. We decided to cut the range of $\lambda_c \pm 2.5 \;\text{\AA}$ around the central wavelength ($\lambda_c$) of each calcium line, so we removed three $5\;\text{\AA}$-wide ranges (that is 18 pixels). The width of the cut was determined by visual inspection of the most active spectra where the emission peak is the most prominent (well above the continuum level) and hence the widest (see Fig. \ref{fig.example_spectrum}; the omitted wavelength range is marked with bands). Most emission features are typically shallower and affect only cores of calcium lines. Although all three cuts intervene far into the wings of Ca lines, there are other relatively strong spectral lines in the spectra (mostly Fe~I and Ti~I lines), so that the $\chi^2$ nearest neighbour search can work. Another reason for the relatively wide omitted range are quite numerous cases of moderate velocity shifts of the emission components.

The distance between two spectra was characterized as 
\begin{equation}
\chi^2=\sum_i{\left( f_\mathrm{a}^i-f_\mathrm{n}^i\right)^2}
\label{eq.chi2}
\end{equation}
where $f_\mathrm{a}^i$ and $f_\mathrm{n}^i$ denote flux values for the $i$~-th wavelength bin of the active and normal spectrum, respectively.

\begin{figure*}
\includegraphics[width=\textwidth]{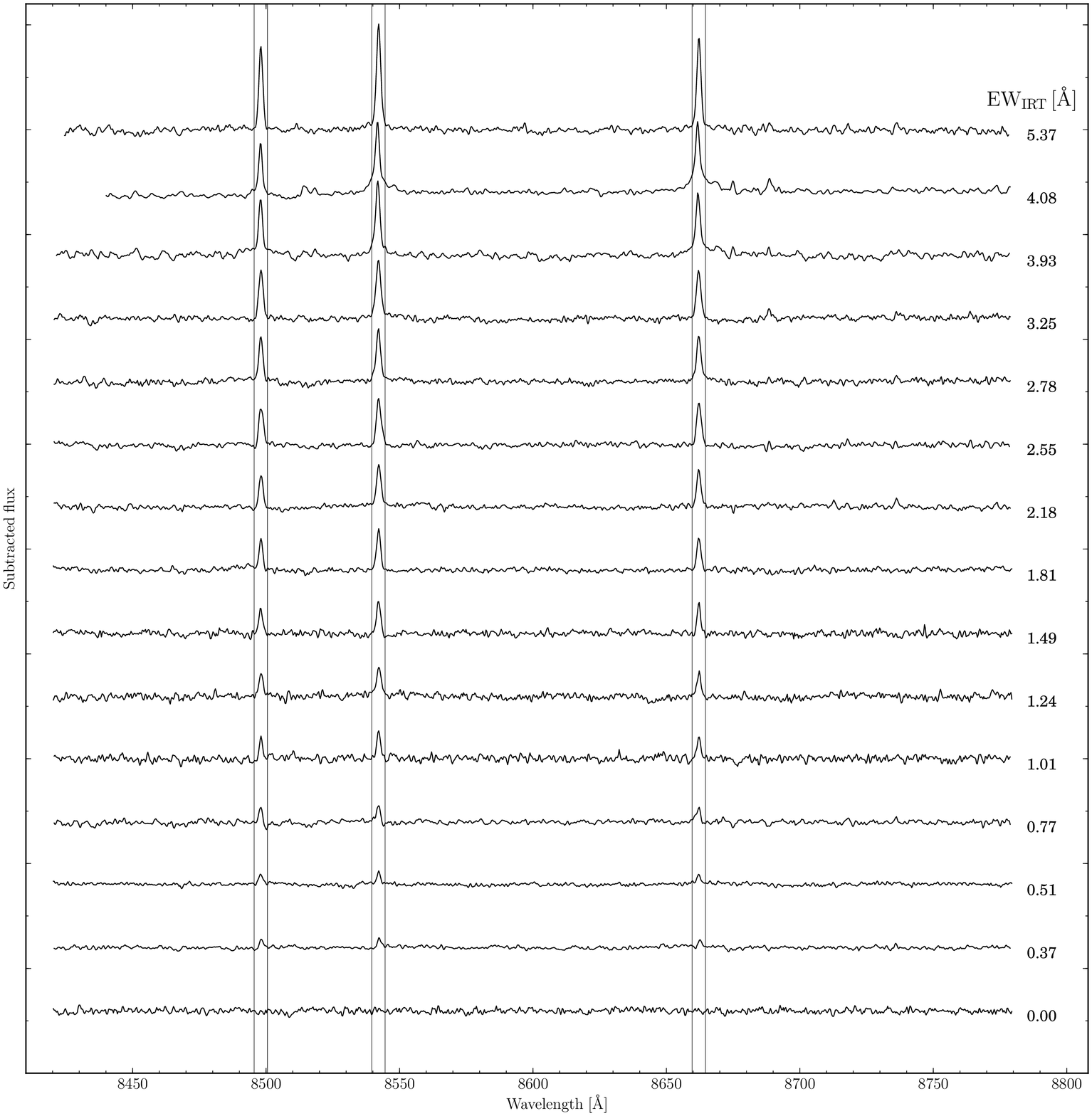}
\caption{Sequence of emission component spectra arranged from an inactive star at the bottom to one of the most active ones on the top. Labels on the right give the $\mathrm{EW_{IRT}}$ value. The three bands indicate wavelength integration ranges used to derive the equivalent widths of the emission components.}
\label{fig.example_spectra}
\end{figure*}

Equation \ref{eq.chi2} assumes both spectra share a common wavelength scale. This way all the spectra in both databases were linearly interpolated to the same wavelength bins. We chose the same interpolation wavelength range and step size as was used in the LLE computation: $8422 \text{\AA} \le \lambda \le 8780 \text{\AA}$ with 1200 bins linear in wavelength, i.e.\ with $\sim 0.3\;\text{\AA}$ spacing. This represents a moderate oversampling over the observed $\sim 0.37\;\text{\AA}$ sampling. Lower and upper interpolation wavelength limits were chosen as typical observed spectrum limits. However, a small fraction of RAVE spectra exists where the interpolation wavelength range is not completely covered. Due to this fact we compare only the intersection of both spectra each time.

All RAVE spectra are continuum normalized. However, the normalization might in some cases not be ideal and continuum levels of the two spectra may differ. All normal spectra in the database were multiplied so that their median level matched the median of an active spectrum. All medians were derived ignoring the windows around calcium lines. This was done for each active spectrum from the database.

A basic underlying assumption of the $\chi^2$ comparison is that all RAVE spectra are shifted to a zero heliocentric radial velocity. Since the selection of all spectra from a normal database is based on the results of the LLE (where the same assumption holds true), their velocity values are highly reliable. However, this may not be the case for active spectra because emission components of the calcium line could affect the parameter determination and cause radial velocity jitter \citep{2010A&A...520A..79M}. Hence the nearest neighbour search was performed iteratively: in each iteration the best matching template was found and the active spectrum was shifted in velocity space with respect to the template. Due to missing calcium lines only a part of the whole spectrum ($8668 \;\text{\AA} \le \lambda \le 8780 \;\text{\AA}$) with relatively strong Fe~I spectral lines ($8674.75, \;8688.63\;\text{\AA}$) was used to compute cross-correlation functions and to derive the radial velocities. 
Spectra were shifted iteratively until the last correction decreased below $1\;\mathrm{km\,s^{-1}}$ (that is 10~\% of pixel width, which is $10.5\;\mathrm{km\,s^{-1}}$). The errors of the computed radial velocities are based on the dispersion of data points of the cross-correlation function 
(standard deviation, denoted by $\sigma$) around a fitted parabola and correspond to the half-width of the parabola at $1\,\sigma$ level below the maximum of the function. If the Fe lines are too weak, the cross-correlation function peak is wide and brings larger errors.

The overall procedure for each active spectrum is as follows:
\begin{enumerate}
\item Interpolate all spectra (both normal and active) to match the same wavelength bins.\label{item.interpolation}
\item Cut calcium lines from all spectra (both active and normal) at $\lambda_c \pm 2.5\;\text{\AA}$. \label{item.cut}
\item Renormalize spectra based on their median value comparison.
\item Apply a nearest neighbours search by $\chi^2$. \label{item.knn}
\item Compute a potential radial velocity correction of the spectrum - take only selected part of spectrum.
\item If a RV correction is applied, repeat the loop from step \ref{item.cut}.
\item If the RV correction is below $1\;\mathrm{km\,s^{-1}}$, exit the loop.
\end{enumerate}

Step \ref{item.interpolation} was done only once at the beginning to save time and was not repeated each time in a loop due to its independence of any variable parameters.

\subsubsection{New stellar parameters}\label{sec.parameters}
\begin{figure}
\includegraphics[width=\columnwidth]{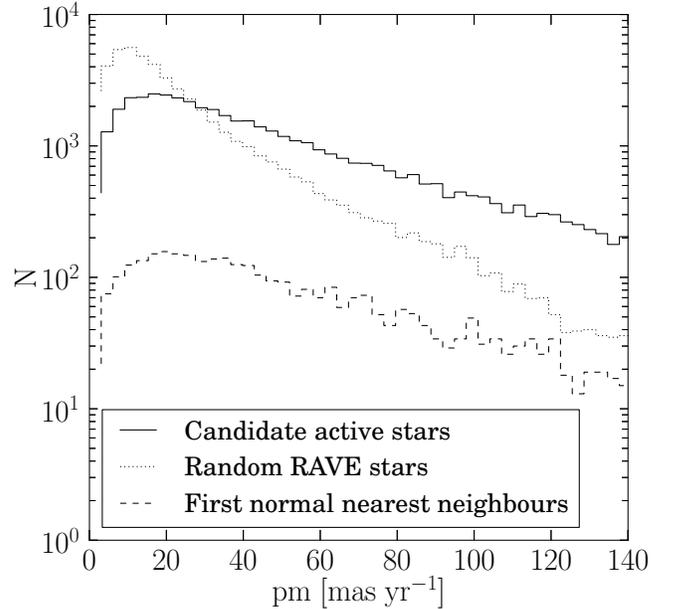}
\caption{Comparison of proper motions of candidate active stars (solid line, $\sim 50,000$ stars) with their nearest normal neighbours (dashed line) and randomly chosen stars from the overall RAVE sample (dotted line, $\sim 50,000$ stars). The median values of the distributions are $\sim17\;\mathrm{mas\;yr^{-1}}$ for the whole sample and $\sim40\;\mathrm{mas\;yr^{-1}}$ for candidate active stars. The distributions are in the logarithmic scale. The comparison supports the assumption that the active sample consists mainly of dwarf stars.}
\label{fig.proper_motion}
\end{figure}

Each candidate active spectrum was assigned with the new stellar parameters (effective temperature, metallicity and gravity). Although the signs of chromospheric activity were already observed in giants as well (see for example \citealp{1999ApJ...520..751D}), we assumed that all normal and candidate active spectra are dwarfs with $\log{g}>3.5$ and hence they share similar values of surface gravity due to the fact that this is a very weak and unbound parameter in the RAVE sample.
The partial support for our assumption is Figure \ref{fig.proper_motion} that compares proper motions of three sets of data: candidate active stars ($\sim 50,000$), their nearest normal neighbours ($\sim 3,600$ because some candidate active stars share the same normal neighbours) and randomly chosen stars ($\sim 50,000$) from the whole RAVE sample. It is clearly seen that candidate active stars and their nearest neighbours (these are normal stars with $\log{g}>3.5$) show a similar shape of the distribution while the overall sample significantly differs.
Candidate active sample and their neighbours contain larger fraction of stars with larger proper motions than the overall sample (the median values of the distributions are $\sim17\;\mathrm{mas\;yr^{-1}}$ for the whole sample and $\sim40\;\mathrm{mas\;yr^{-1}}$ for candidate active stars). Approximately half of the whole RAVE sample is represented by giant stars. Since a fraction of the detected giants reaches larger distances due to their larger luminosities compared with dwarfs, these stars lower the median value of the distribution of the proper motion because the average measured proper motion of more distant star is smaller. Larger median proper motion for candidate active stars together with the distribution similar to the part of the normal database indicate that the candidate active sample is dominated by dwarfs and the number of giants is small.

New stellar parameters were defined as the arithmetic mean of the original parameters (from RAVE pipeline) of its first five nearest normal neighbours. Scatter of these new parameters was estimated as their standard deviation where the internal uncertainty of each parameter was ignored.

All values of the new parameters of course fall within the normal database range. Figure \ref{fig.database_parameters} shows the distribution of the temperature and metallicity for both normal and active spectra (the newly defined parameters). The temperature distribution of active spectra is centered around $5000\;\mathrm{K}$ and the average is lower than in the normal database.
The distribution of the active objects' metallicities matches the distribution of the normal database.

Nonzero radial velocity shifts of both normal and active spectra were investigated (normal spectra were not shifted; only a potential shift was calculated for reference).
The mean value of shifts for the normal database is zero with a standard deviation of $2.5\;\mathrm{km\,s^{-1}}$. Corrections were applied on $\sim 61~\%$ of all active spectra. The distribution of our radial velocity correction for active spectra has a zero mean value with a standard deviation of $19\;\mathrm{km\,s^{-1}}$. The amplitude of the radial velocity correction for 1812 spectra (that is 6~\% of all nonzero corrections) is larger than $1\;\sigma_{\mathrm{RV}}$ and 173 spectra cross $3\;\sigma_{\mathrm{RV}}$ level. Here $\;\sigma_{\mathrm{RV}}$ is the uncertainty on the derived radial velocity. Median value of all uncertainties is $5.8\;\mathrm{km\,s^{-1}}$. 5~\% of the stars with nonzero corrections have velocities $\ge 5.25\;\mathrm{km\,s^{-1}}$ (half of the pixel). 99~\% of nonzero RV corrections for active stars fall within $\pm 10.5\;\mathrm{km\,s^{-1}}$ (1 pixel) and there are 101 stars with the values greater than $\pm 21\;\mathrm{km\,s^{-1}}$ (2 pixels).

However, there are $\sim20$ very active stars with a displacement of several hundreds of $\mathrm{km\,s^{-1}}$. The reason is that the most prominent features in these spectra are calcium emission lines while the RAVE radial velocity pipeline was designed for absorption spectra and applied wrong and very large shifts.
Radial velocity corrections allowed us to discover the most active spectra in our sample.
RV corrections are given in Table \ref{sec.catalogue} in the catalogue.

\subsection{Subtracted spectra}
The chromospheric emission flux was extracted by a subtraction method. The nearest normal neighbour was subtracted from an active spectrum to remove the photospheric contribution to the total flux. The subtracted spectrum contains only the emission flux of the object, the rest oscillates around zero due to the noise. The exception are wavelengths of calcium line cores with chromospheric emission. In the case of a nearly quiet activity state the emission peaks are buried in the noise. With an increased activity they become more and more apparent (see Fig. \ref{fig.example_spectra}).

$58\;\%$ of the subtracted spectra have the highest pixel value among the 18 pixels within the strongest calcium line (at $8542\;\text{\AA}$) crossing the $2\,\sigma_S$ level and $15\;\%$ of them cross the $5\,\sigma_S$ level. Here $\sigma_S$ is the standard deviation of the subtracted spectrum outside the calcium line ranges.

\begin{figure*}
\includegraphics[width=\textwidth]{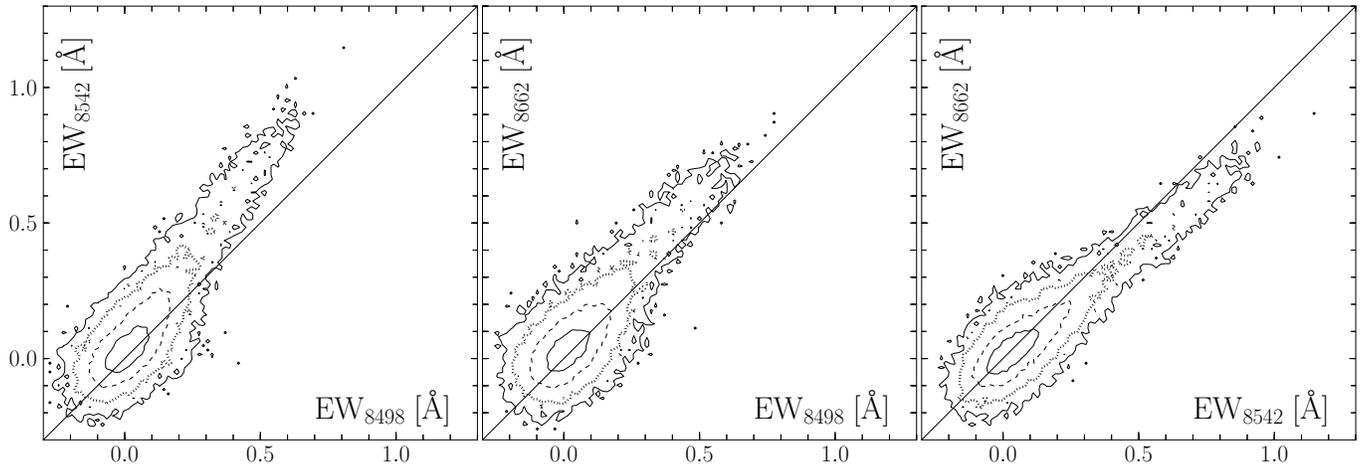}
\caption{Comparison of emission flux equivalent widths for all three calcium lines. Different linestyles indicate the number of objects per bin on a particular contour. Bin size is 0.017~\AA. The scale is logarithmic. Contour values are 2 stars for the outer solid line, 11 for the dotted, 55 for the dashed and 181 stars for the inner solid line.}
\label{fig.line_correlations}
\end{figure*}

\section{Activity}\label{sec.activity}

In this paper the activity was computed for each of the three calcium lines first as an equivalent width $\mathrm{EW_{\lambda}}$ in the subtracted spectrum using the same wavelength range as for the calcium cuts (width of $5\;\text{\AA}$, that is 18 pixels for each line). The computation was done 5 times, by subtracting in turn each of the five nearest neighbours from the normal database.
It turned out that an average $\mathrm{EW_\lambda}$ computed with the first 5 nearest neighbours was not significantly different from an average using 10, 15 or 20 nearest neighbours, so we kept the value derived using 5 nearest neighbours. The stability of $\mathrm{EW_{\lambda}}$ confirms the assumption that our normal database completely and sufficiently covers the whole parameter space. However, it may happen in rare cases that the number of nearest neighbours that adequately fit the active spectrum is less than five and that their successors have slightly different spectra. In that case we still take first five neighbours, but the errors (the scatter) of the equivalent widths, which are estimated as the standard deviation of activity levels within that 5 neighbours, are higher. The typical error on the equivalent width of the emission component of each calcium line is $0.05\;\text{\AA}$.

The equivalent width of the chromospheric flux for each particular line ranges from $\sim -0.2\;\text{\AA}$ to $\sim 1\;\text{\AA}$, or even more in the case of the most active stars. The reason for negative values is a template mismatch in the case of marginally active or inactive stars. Another possibility is that the chromospheric flux is dominated by noise. Yet another reason could be that there are no non-active stars even in the normal database since every star shows at least minimal signs of activity, but the detection of such low levels of activity is well below our limits in the RAVE database.

Fig. \ref{fig.line_correlations} shows correlations of activity between all three lines. The most obvious characteristic of all three plots is the distribution of activity levels which are concentrated between $\sim -0.1\;\text{\AA}$ and $\sim 0.25\;\text{\AA}$. Such a result is expected because of the selection criteria. Our active database consists of all stars that have at least one active neighbour within the first six neighbours (LLE based classification). This is a relatively weak criterion as we wish to include marginally active candidate stars as well. However, stars with no statistically significant activity can be discarded later for the purposes of further analyses.

The second feature visible on the diagrams are tightly correlated strengths of lines compared to each other. One can easily see that the $8542\;\text{\AA}$ line is the strongest one and the $8498\;\text{\AA}$ line the weakest one. A good correlation is another test for consistency of the results and excludes the possibility that the emission signal comes from another source like cosmic rays or similar errors. 
The individual panels of Fig. \ref{fig.line_correlations} compare two of the lines.
The error estimation for each calcium line was computed as a scatter (a standard deviation) among the equivalent widths obtained by subtracting the first five nearest normal neighbours ($0.05\;\text{\AA}$ in all three cases). Similarly, the dispersion around zero for normal stars for each line is $0.06\;\text{\AA}$.

\begin{figure*}
\includegraphics[width=\textwidth]{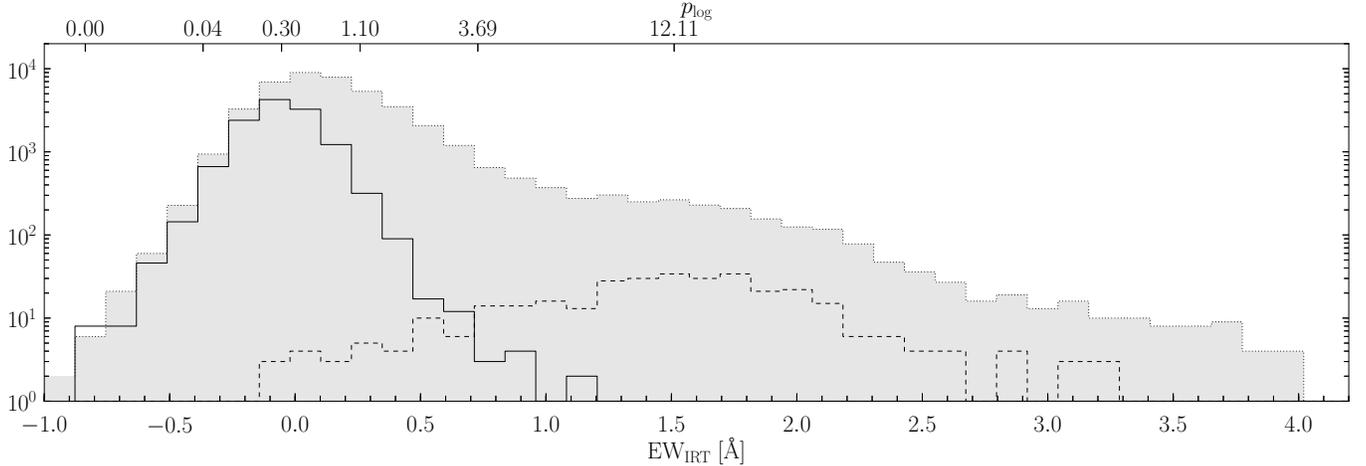}
\caption{Distribution of $\mathrm{EW_{IRT}}$ for active stars (grey area), normal stars (which are assumed to be inactive; solid line) and pre-main sequence stars (selection is based on Simbad classification of RAVE stars; dashed line). The scale is logarithmic. $p_{\mathrm{log}}$ is a measure for the probability that a star with given $\mathrm{EW_{IRT}}$ differs from an inactive spectrum. $p_{\mathrm{log}}$ values correspond to 5 and 2~$\sigma$ below zero and 2, 5 and 10~$\sigma$ above zero. 0.3 marks zero.}
\label{fig.activity_histogram}
\end{figure*}

To reduce the dimensionality of the problem and to improve significance of the results we introduce the sum of equivalent widths 
\begin{equation}
\mathrm{EW_{IRT}=EW_{8498}+EW_{8542}+EW_{8662}} \nonumber
\end{equation}
for all three calcium lines for each star. The overall distribution of activity ($\mathrm{EW_{IRT}}$) is shown in Fig. \ref{fig.activity_histogram} and clearly shows a bimodal distribution with two peaks at $\sim 0.1$ and $\sim 1.5\;\text{\AA}$. For reference, the activity of normal stars was derived with the same method (we took 2nd-6th nearest neighbours since the first was the spectrum itself). As expected the distribution of values in the normal database is Gaussian-shaped around $-0.05\;\text{\AA}$ with a standard deviation of $\sigma = 0.16\;\text{\AA}$.
This value is adopted as a general uncertainty in any $\mathrm{EW_{IRT}}$ value for both normal and active stars. The second reference sample are pre-main sequence stars - RAVE stars that are classified as such in the Simbad database. This distribution overlaps with the second peak of active stars and perhaps implies a discovery of pre-main sequence stars in the RAVE survey. This topic will be explored in our next papers.

Another source of uncertainty could be the time variability of active spectra. The HK Project contains 40 years of data that showed the variability of activity. Our results may reflect the activity level at the time of observation, but this can change with time.

The abscisa axis along the top axis of Fig. \ref{fig.activity_histogram} shows the measure of probability $p_{\mathrm{log}}$ that a star with a given $\mathrm{EW_{IRT}}$ and $\sigma$ is more active than a similar inactive normal star. If we have two Gaussian distributions with $(\mathrm{EW_{IRT}^0}, \; \sigma_0)$ and $(\mathrm{EW_{IRT}^1}, \; \sigma_1)$ where the equivalent width denotes the mean value and $\sigma^2$ the variance of the distribution, and we randomly pick one sample from each distribution, the probability that $\mathrm{EW_{IRT}^1}>\mathrm{EW_{IRT}^0}$ is 
\begin{equation}
P(\mathrm{EW_{IRT}^1}>\mathrm{EW_{IRT}^0}) = \frac{1}{2} \left[ 1 + \mathrm{erf}{\left( \frac{\mathrm{EW_{IRT}^1}-\mathrm{EW_{IRT}^0}}{\sqrt{2(\sigma_0^2+\sigma_1^2)}} \right)} \right] \nonumber
\end{equation}
(for derivation see \citealp{2011AJ....141..200M}).
Because the error function $\mathrm{erf}(z)$ reaches values very close to 1 at $z=2$, we introduced
\begin{equation}
p_{\mathrm{log}}=-\log_{10}(1-P) \nonumber
\end{equation}
to make differences more distinguishable. However, due to a limited computational precision we had to stop at $P\approx1-10^{-13}$ where $p_{\mathrm{log}}\approx 12$. If stars have exactly the same equivalent width, then $P=1/2$ and $p_{\mathrm{log}}=0.3$. The higher the $P$, the higher is $p_{\mathrm{log}}$. Values on the plot were computed with $\sigma_0=\sigma_1=0.16\;\text{\AA}$ and $\mathrm{EW_{IRT}^0}=-0.05\;\text{\AA}$ (this is the mean value of the distribution of normal stars).

46~\% of stars in our active database have $\mathrm{EW_{IRT}}$ below $1\;\sigma$ and 54~\% above this value. Further, there are 32~\% of stars with $\mathrm{EW_{IRT}}>2\sigma$ (that is $\sim$~14,200 stars, $p_{\mathrm{log}}=1.10$), 19~\% with $\mathrm{EW_{IRT}}>3\sigma$ ($p_{\mathrm{log}}=1.77$) and 12~\% with $\mathrm{EW_{IRT}}>4\sigma$ ($\sim$~5,200 stars, $p_{\mathrm{log}}=2.63$).

\section{Discussion and conclusions}\label{sec.discussion}
This paper quantitatively describes our catalogue of candidate active RAVE stars. It significantly enlarges previously known samples. It will become public and be part of the next data release. Our catalogue is based on the LLE classification of stellar spectra (\citealp{2012ApJS..200...14M}) and contains all spectra that have at least one chromospherically active nearest neighbour within its first 6 neighbours. However, since the other five neighbours can be of any type and because we analysed all stars with affected calcium lines with no respect to the origin of the additional emission flux (also, besides the possibility of problems with continuum normalization in some very rare cases with the underestimated $S/N$ a large noise may fill the calcium lines and mimic the emission), our results give candidates for active stars, should be treated with caution and might not be error free. The estimation for activity levels is independent of the RAVE pipeline version because our method is based on a direct comparison of fluxes and omits the use of stellar parameters.

A large database of normal spectra (high S/N single-lined main-sequence dwarfs without any peculiarities) was used to apply a spectral subtraction technique on candidate active stars in order to extract the equivalent widths of emission flux in Ca~II IRT lines. Although all spectra were previously shifted to zero heliocentric radial velocity with the RAVE pipeline, it turned out that moderate and very active candidate spectra still suffered from radial velocity displacements due to their shallower or even emission-like calcium lines that could not be treated properly in the pipeline. Therefore, radial velocity corrections had to be applied on 61~\% of candidate active spectra ($\ge 5.25\;\mathrm{km\,s^{-1}}$ for 5~\% of candidate active stars with nonzero correction). For more than 99~\% of stars with nonzero correction the amplitude of shift was up to $10.5\;\mathrm{km\,s^{-1}}$. However, there were $\sim20$ exceptional cases of very active candidate stars with an emission-like spectrum and with displacement of several hundreds of $\mathrm{km\,s^{-1}}$ due to the improper previous treatment in the RAVE pipeline caused by the strong emission nature of the spectra.

Our analysis shows a good correlation of activity between all three lines (Fig. \ref{fig.line_correlations}). A distribution of stellar activity rates $\mathrm{EW_{IRT}}$ has a clear bimodal shape where the most abundant peak matches the inactive sample and the most active peak arises at activity levels of pre-main sequence stars (Fig. \ref{fig.activity_histogram}). The most active candidate stars reach equivalent widths of $\sim 5\;\text{\AA}$. While the majority of the sample comprises of shallower calcium lines, there appear clear additional emission peaks within the lines in moderate and very active candidate stars. Some of them even show Doppler shifts that will be discussed in our future paper.

\begin{figure}
\includegraphics[width=\columnwidth]{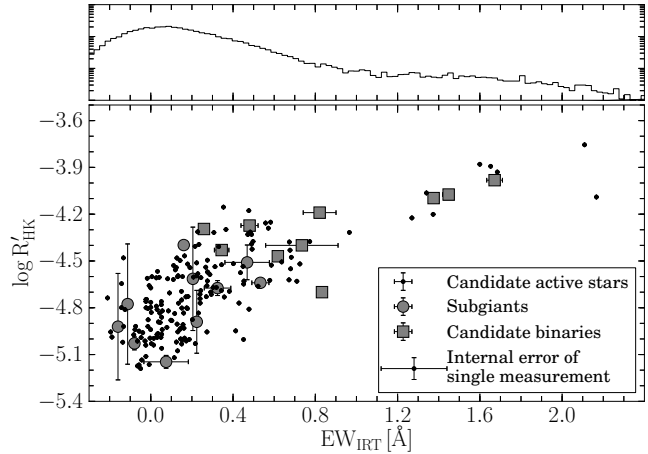}
\caption{$\log{\mathrm{R'_{HK}}}$ versus $\mathrm{EW_{IRT}}$ values for 211 RAVE stars matched with a set of available online catalogues (they can contain different populations of stars because their selection criteria might differ). Some of the stars were measured multiple times by various authors; in this case we plot an average value for the star. The typical error of $\mathrm{EW_{IRT}}$ is $0.16\;\text{\AA}$. Our estimation of the $\log{\mathrm{R'_{HK}}}$ scatter is 0.07. Big circles indicate 'subgiants' - stars that are classified as luminosity class III, IV or IV-V in the Simbad database. Squares are binary candidates found by the LLE classification. The internal error of single measurement marks the internal error for $\mathrm{EW_{IRT}}$ while its ordinate error value is the $\log{\mathrm{R'_{HK}}}$ scatter. The histogram in the top panel shows the distribution of $\mathrm{EW_{IRT}}$ for the whole sample of active RAVE stars in logarithmic scale.}
\label{fig.ewirt_vs_rhk}
\end{figure}

The comparison of our $\mathrm{EW_{IRT}}$ and $\log{\mathrm{R'_{HK}}}$ (for stars from the literature) forms another supplementary table to the catalogue. 543 measurements of 211 matching stars were found in 21 different catalogues from the northern and southern sky: \citealp{1991ApJS...76..383D, 1995ApJ...438..269B, 1996AJ....111..439H, 1998MNRAS.298..332R, 2000A&AS..142..275S, 2002MNRAS.332..759T, 2003AJ....126.2048G, 2004ApJS..152..261W, 2006AJ....132..161G, 2006MNRAS.372..163J, 2007A&A...469..309C, 2007AJ....133..862H, 2007ApJS..171..260L, 2007AJ....133.2524W, 2009A&A...493.1099S, 2010ApJ...725..875I, 2010A&A...514A..97L, 2010A&A...520A..79M, 2011ApJ...734...70A, 2011A&A...531A...8J, 2011PASP..123..412W}. In the case of \citealp{1991ApJS...76..383D, 1995ApJ...438..269B}, \citet{2003AJ....126.2048G} (table 5), \citealp{2007A&A...469..309C, 2010A&A...514A..97L} the S-index was converted to $\log{\mathrm{R'_{HK}}}$ using the relation introduced by \citet{1984ApJ...279..763N}. $B-V$ colours used in the derivation of $\log{\mathrm{R'_{HK}}}$ for stars from \citealp{1991ApJS...76..383D, 2010A&A...514A..97L} were derived using Simbad values. We also checked \citet{2010ApJ...720.1569K} but found no matching stars with the RAVE database.
Figure \ref{fig.ewirt_vs_rhk} shows correlation between $\mathrm{EW_{IRT}}$ and $\log{\mathrm{R'_{HK}}}$. 
There exists a strong correlation between the two. However, most stars are scattered around a less active area and therefore we assume that this sample is subjected to relatively high intrinsic variability and measurement errors. This sample consists of single measurements and of stars with multiple observations (either in time or in more than one catalogue); in these cases the average values are plotted. The typical scatter of $\log{\mathrm{R'_{HK}}}$ in this figure is 0.07. 
This is the median value of standard deviations of multiple observations. Uncertainties on $\log{\mathrm{R'_{HK}}}$ in \citet{1996AJ....111..439H} are comparable to our estimation from multiple observations. It seems that the sample of 211 matching catalogue stars contains mostly stars with a lower activity rate than RAVE (around $\log{\mathrm{R'_{HK}}}=-5.1$). However, the most active stars from various authors reached $\log{\mathrm{R'_{HK}}}\approx -3.8$ with the second (most active) peak at $\log{\mathrm{R'_{HK}}}\approx -4.5$. Figure \ref{fig.ewirt_vs_rhk} shows that $\log{\mathrm{R'_{HK}}} \sim -5$ is undetectable in the RAVE survey by our measurement technique. Most of the stars in our catalogue therefore have only a marginal detection of activity. But on the other hand there are numerous very active candidate objects discovered worth of further detailed analysis. Although various authors report a gap in the bimodal $\log{\mathrm{R'_{HK}}}$ distribution at $\log{\mathrm{R'_{HK}}} \approx -4.75$ (a rough estimate; the gap is called the Vaughan-Preston gap), we suspect that the possible existence of such a gap in our distribution is difficult to detect.

Based on the 58 age estimations of 46 matched stars from three of the catalogues (\citealp{2004ApJS..152..261W, 2010ApJ...725..875I, 2011ApJ...734...70A}) the ages of stars vary from $100\;\mathrm{Myr}$ at relatively high activity levels ($\log{\mathrm{R'_{HK}}} \sim -4.3$) to $10\;\mathrm{Gyr}$  at $\log{\mathrm{R'_{HK}}} \sim -5.2$.

\begin{figure}
\includegraphics[width=\columnwidth]{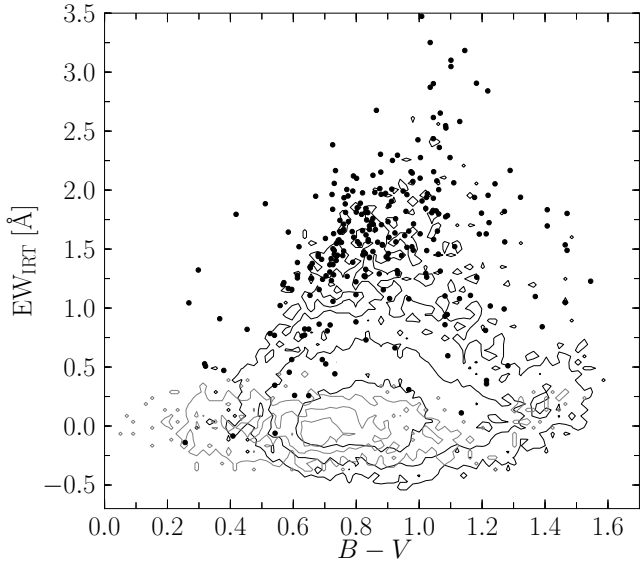}
\caption{Color dependency ($B-V$) of $\mathrm{EW_{IRT}}$. Grey contours correspond to the normal database (2, 7 and 20 stars per bin; note that a fraction of normal stars lacks of $B-V$ data), black lines indicate contours for active stars (2, 11 and 55 stars per bin). RAVE pre-main sequence stars (Simbad classification) are added for reference (black dots). The bin size is 0.02 in color and $0.05\;\text{\AA}$ in $\mathrm{EW_{IRT}}$.}
\label{fig.b-v_ew}
\end{figure}

\begin{figure*}
\includegraphics[width=\textwidth]{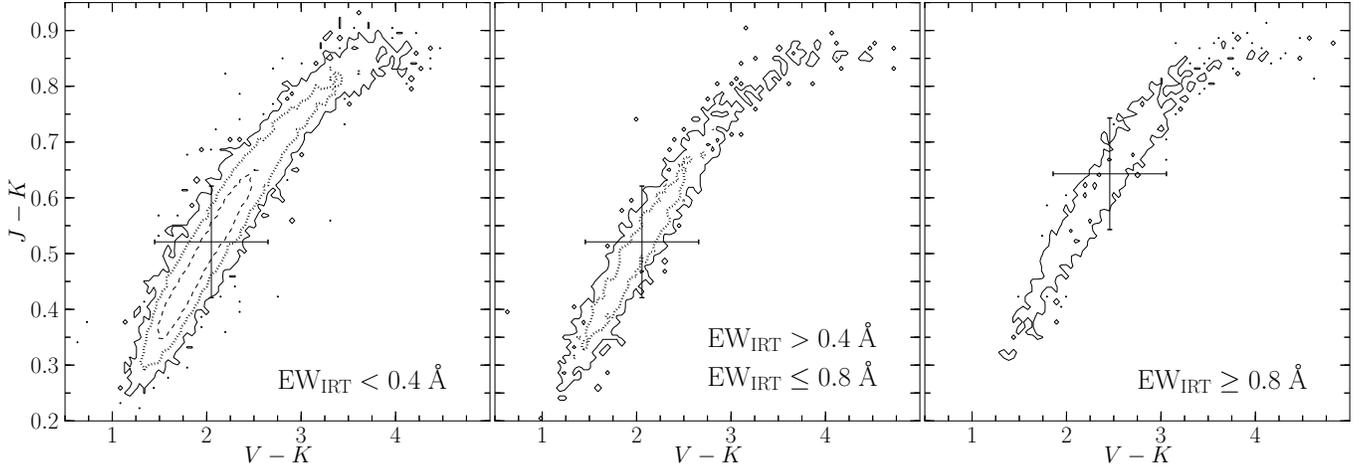}
\caption{Color-color diagram (J-K versus V-K) for candidate active stars. J and K are 2MASS magnitudes, V is from the APASS catalogue. The whole candidate active dataset is divided into three subsets according to their $\mathrm{EW_{IRT}}$. The median of the distribution (cross on each diagram) moves from $(J-K, V-K)=(0.52, 2.05)$ for stars with a weak activity to a much redder value of $(0.64, 2.46)$ for the most active candidate stars. The bin size is 0.05 for $V-K$ and 0.01 for $J-K$. Contour levels are number of stars per bin: 2, 11 and 67.}
\label{fig.colorcolordiagram}
\end{figure*}

This catalogue gives the opportunity to further study the possible parameter dependence of activity which will be explored in our next paper. 
Line assymetries could be studied and since the RAVE sample is large, a discovery of peculiars and particularly interesting stars among the active sample is possible. 
\citet{siviero_peculiars_2013} are working on the ongoing observational project of RAVE peculiars (at Asiago Observatory), including the chromospherically active stars, that will provide the additional observations.
Another topic to be discussed in our ongoing project is dependence of activity on stellar populations and the photospheric physics. As an example, Figures \ref{fig.b-v_ew} and \ref{fig.colorcolordiagram} show dependence of activity on color ($B-V$) and color-color diagram where the most active candidate stars move to the reddest parts of the plot. In Figure \ref{fig.colorcolordiagram} we used the APASS magnitudes as given in the DR6 public data release \citet{2012JAVSO..40..430H}. 
Furthermore, \citet{2006AJ....132..161G} discovered an absence of the Vaughan-Preston gap for $[M/H]<-0.2$ and that implies a question of metallicity dependence of activity level that could be investigated.

It is well known that chromospheric emission anti-correlates with stellar ages (e.g.\  \citealp{2010ARA&A..48..581S}). We plan to investigate and use of RAVE stars which are members of stellar clusters (\citealp{conrad_2013_1}) to calibrate the RAVE's $\mathrm{EW_{IRT}}$ with age and to explore the spatial distribution, the distribution with distances from Galactic plane as well as the kinematics of candidate active stars in our Galaxy.

Since the RAVE catalogue of candidate active stars consists of a relatively large number of objects it will be a suitable and convenient database to be compared with the results of upcoming Gaia space mission (it covers Ca~II IRT; \citealp{2004MNRAS.354.1223K, 2005MNRAS.359.1306W}) and Hermes-Galah (which covers $\mathrm{H\alpha}$ and $\mathrm{H\beta}$; \citealp{2012ASPC..458..421Z}).

\section*{Acknowledgements}
We would like to acknowledge the very relevant and helpful comments of an anoymous referee which substantially improved the quality and presentation of the paper.

Funding for RAVE has been provided by: the Anglo-Australian Observatory; the Leibniz-Institut fuer Astrophysik Potsdam; the Australian National University; the Australian Research Council; the French National Research Agency; the German Research foundation; the Istituto Nazionale di Astrofisica at Padova; The Johns Hopkins University; the National Science Foundation of the USA (AST-0908326); the W.M. Keck foundation; the Macquarie University; the Netherlands Research School for Astronomy; the Natural Sciences and Engineering Research Council of Canada; the Slovenian Research Agency; Center of Excellence Space.si; the Swiss National Science Foundation; the Science \& Technology Facilities Council of the UK; Opticon; Strasbourg Observatory; and the Universities of Groningen, Heidelberg and Sydney. The RAVE web site is at \href{http://www.rave-survey.org}{http://www.rave-survey.org}.

This research has made use of the SIMBAD database and the VizieR catalogue access tool, operated at CDS, Strasbourg, France.

GM acknowledges support by the NASA ADAP grant number NNX13AF32G.

\newpage
\appendix
This section is a description of the catalogue of candidate active RAVE stars and includes three tables of newly derived parameters. Appendix \ref{sec.catalogue} is the main catalogue of this paper and contains the $\mathrm{EW_{IRT}}$ of 53,347 stars.

The table in Appendix \ref{sec.rhkew} gives both RAVE $\mathrm{EW_{IRT}}$ and $\log{\mathrm{R'_{HK}}}$ available from the literature to compare the results (211 stars). Both tables are based on RAVE data obtained between 2004 April 8 and 2012 October 20. Since RAVE is an ongoing project, new data will be added in the future.

\section{A. The catalogue}\label{sec.catalogue}
This is the description of the output catalogue of $\sim$~44,000 RAVE candidate active stars that will be accessible through Vizier. Among usual and obvious values it contains newly-estimated stellar parameters (Teff\_K2, Logg\_K2, Met\_N\_K2) which are the arithmetic mean of the parameters of a star's first five nearest normal neighbours. The scatter of such estimates is derived as the standard deviation of these 5 values. RVshift is our correction to the original value (only values that exceed $3\;\sigma_{RV}$ due to the $S/N$ limitations) and RV is the corrected radial velocity. EW8498, EW8542 and EW8662 are average values for each line (average over 5 neighbours), their errors are derived in the same way as in the previous case. The same holds true for EWirt which denotes $\mathrm{EW_{IRT}}$. 

\bigskip
\setlength{\LTpre}{0pt}
\setlength{\LTpost}{0pt}
\begin{longtable}{l l l p{7cm}}
\hline
Acronym & Type & Unit & Description \\
\hline
rave\_obs\_id & string & ... & Targed designation \\
RAdeg & float & deg & Right ascension (J2000.0) \\
DEdeg & float & deg & Declination (J2000.0) \\
STN\_K & float & ... & S/N \\
EWirt & float & \AA & $\mathrm{EW_{IRT}}$ \\
eEWirt & float &\AA & $\mathrm{EW_{IRT}}$ error \\
nnid00 & string & ... & Target designation of star's 1st normal neighbour \\
nnid01 & string & ... & Target designation of star's 2nd normal neighbour \\
nnid02 & string & ... & Target designation of star's 3rd normal neighbour \\
nnid03 & string & ... & Target designation of star's 4th normal neighbour \\
nnid04 & string & ... & Target designation of star's 5th normal neighbour \\
Teff\_K2 & float & K & New effective temperature \\
eTeff\_K2 & float & K & Error of new effective temperature \\
Logg\_K2 & float & dex & New $\log{g}$ \\
eLogg\_K2 & float & dex & Error of new $\log{g}$\\
Met\_N\_K2 & float & dex & New $[M/H]$ \\
eMet\_N\_K2 & float & dex & Error of new $[M/H]$ \\
EW8498 & float & \AA & $\mathrm{EW_{8498}}$ \\
eEW8498 & float & \AA & $\mathrm{EW_{8498}}$ error \\
EW8542 & float & \AA & $\mathrm{EW_{8542}}$ \\
eEW8542 & float & \AA & $\mathrm{EW_{8542}}$ error\\
EW8662 & float & \AA & $\mathrm{EW_{8662}}$ \\
eEW8662 & float & \AA & $\mathrm{EW_{8662}}$ error \\
RVshift & float & $\mathrm{km\,s^{-1}}$ & Our RV correction due to activity \\
eRVshift & float & $\mathrm{km\,s^{-1}}$ & Error of RV correction due to activity \\
RV & float & $\mathrm{km\,s^{-1}}$ & Total heliocentric RV \\
eRV & float & $\mathrm{km\,s^{-1}}$ & Error of total heliocentric RV
\label{tab.catalogue_description}
\end{longtable}

\section{B. Comparison of \texorpdfstring{$\log{\mathrm{R'_{HK}}}$}{R'\_HK} and \texorpdfstring{$\mathrm{EW_{IRT}}$}{EW\_IRT}}\label{sec.rhkew}
This table is a list of 543 measurements of 211 RAVE stars that were found in the literature and observed by other authors in the Ca~II H~\&~K lines. It contains $\log{\mathrm{R'_{HK}}}$ and $\mathrm{EW_{IRT}}$ values that were compared in Fig. \ref{fig.ewirt_vs_rhk}. S-indices from \citealp{1995ApJ...438..269B, 2007A&A...469..309C} were converted to $\log{\mathrm{R'_{HK}}}$ using $B-V$ from the same catalogues, while colours for stars from \citet{1991ApJS...76..383D, 2010A&A...514A..97L} and table 5 from \citet{2003AJ....126.2048G} were obtained with the Simbad database. LogAge is the logarithm of age of the star (in years) where available (\citealp{2004ApJS..152..261W, 2010ApJ...725..875I, 2011ApJ...734...70A}). If the star is found in catalogues multiple times, we list the average value (average over time or/and different catalogues). The other parameters are a subset of the whole list of parameters from the main catalogue as described in Appendix \ref{sec.catalogue}.

\bigskip
\begin{longtable}{l l l p{7cm}}
\hline
Acronym & Type & Unit & Description \\
\hline
rave\_obs\_id & string & ... & Targed designation \\
RAdeg & float & deg & Right ascension (J2000.0) \\
DEdeg & float & deg & Declination (J2000.0) \\
STN\_K & float & ... & S/N \\
EWirt & float & \AA & $\mathrm{EW_{IRT}}$ \\
eEWirt & float &\AA & $\mathrm{EW_{IRT}}$ error \\
R'HK & float & ... & $\log{\mathrm{R'_{HK}}}$ \\
eR'HK & float & ... & Error of $\log{\mathrm{R'_{HK}}}$\\
logAge & float & ... & $\log_{10}{\mathrm{age}}$ \\
Ref & string & ... & $\log{\mathrm{R'_{HK}}}$ reference \\
RefA & string & ... & Age reference \\
fLLE & char & ... & Average flag from LLE classification \\
Teff\_K2 & float & K & New effective temperature \\
eTeff\_K2 & float & K & Error of new effective temperature \\
Logg\_K2 & float & dex & New $\log{g}$ \\
eLogg\_K2 & float & dex & Error of new $\log{g}$\\
Met\_N\_K2 & float & dex & New $[M/H]$ \\
eMet\_N\_K2 & float & dex & Error of new $[M/H]$ \\
\end{longtable}

\bibliographystyle{apj}
\bibliography{bibliography}
\end{document}